\documentclass[12pt,preprint]{aastex}

\shorttitle{Discovery of a massive SCUBA core}
\shortauthors{Wu et al.}

\begin{document}

\newcommand{\massshort}{1428}
\newcommand{\masslong}{1678}


\title{Discovery of a massive SCUBA core with both inflow and outflow motions}


\author{Yuefang Wu\altaffilmark{1}, Ming Zhu\altaffilmark{2}, Yue Wei\altaffilmark{1},
Dandan Xu\altaffilmark{1},  Qizhou Zhang\altaffilmark{3} and Jason D.
Fiege\altaffilmark{4}}


\altaffiltext{1}{Astronomy Department,
                Peking University, Beijing 100871, China, yfwu@bac.pku.edu.cn}
\altaffiltext{2}{Joint Astronomy Centre/National Research Council of Canada, Herzberg
Institute of Astrophysics, 5071 West Saanich Road, Victoria, BC
V9E 2E7, Canada}
\altaffiltext{3}{Harvard-Smithsonian Center for
Astrophysics, 60 garden Street, Cambridge, MA 02138, U.S.A}
\altaffiltext{4}{Department of Physics \& Astronomy,
University of Manitoba,Winnipeg, Manitoba, R3T 2N2, Canada}


\begin{abstract}

  We report the discovery of a massive  SCUBA core with evidence
  of inflow and outflow motions.  This core is detected by SCUBA at both
  450 and 850 $\mu$m. Barely resolved by the telescope beam at 450 $\mu$m, it has a
  size of 10\arcsec, corresponding to 0.28 pc at a distance of
  5.7 kpc. The dust temperature is estimated to be $\leq$ 29 K, the total
  mass is 820 $\rm M_\sun$ and the average density is
  1.1$\times 10^6$ cm$^{-3}$ in a region with a radius of 5\arcsec.
  Follow-up
  spectral line observations, including HCN (3$-$2), HCO$^+$ (3$-$2),
  H$^{13}$CO$^+$ (3$-$2) and C$^{17}$O (2$-$1) reveal a typical blue
  profile which indicates that this core is collapsing. The CO (3$-$2)
  line profile is as broad as 38 km/s, indicating outflow motions in
  this region. This core is approximately 1.5 pc away from the known HII
  region G25.4NW, but there are no obvious radio, IRAS, MSX or Spitzer
  sources associated with it.  We suggest that this core is at a very
  early stage of massive star or cluster formation.

\end{abstract}


\section{Introduction}

Finding massive sources at their earliest stage of star formation
is crucial for the study of high mass star formation. Such objects
are very difficult to find  due to the highly obscured environment
and fast evolution of high mass stars. Various surveys were
carried out in recent years using molecular lines and dust
continuum emission in millimeter and sub-millimeter wavelength
bands (Molinari et al. 1996, 2002; Sridharan et al. 2002; Beuther
et al. 2002). These surveys, based on IRAS selected samples, have
identified a number of precursors of ultracompact (UC) HII regions
or high mass protostellar (HMPO) candidates. However, the
precursors of UC HII regions or HMPOs usually have $L_{\rm
IR}>10^3L_{\sun}$ indicating that the sources are already in a
relatively developed stage of star formation. Recently, Garay et
al. (2004) reported the discovery of four massive and dense cold
cores, which could be the potential sites for future massive star
formation.

This letter reports the discovery of a massive cold core using the
James Clerk Maxwell Telescope (JCMT). This core was first
identified as an ammonia core 18354$-$0649
(R.A. (1950)=$\rm18^h 35^m26.5^s$, DEC. (1950)= $-06^\circ 49
\arcmin 32 \arcsec $) during an ammonia survey of massive star
formation regions using the MPIfR 100 m telescope at Effelsberg
(Wu et al. 2005). Fig.1a shows the NH$_3$ (1,1) grey scales and
(3,3) contours of the IRAS 18355$-$0650 region overlaid on the VLA
6 cm continuum image (Lester et al. 1985). The center of the
NH$_3$ core is about 80\arcsec \, away from the IRAS source, and
is not associated with any radio continuum emission region. We
have made observations using SCUBA at the JCMT as well as a number
of molecular lines at sub-millimeter wavelengths. The
observational details and data analyses are presented in the
following sections.

\section{Observations\label{obs}}

Dust continuum emission of the ammonia core 18354$-$0649 was
mapped simultaneously at both 850 $\rm \mu m$ and 450 $\rm \mu m$
using SCUBA at the JCMT on Sept. 8, 2003. A 64-point jiggle
pattern was used, with a chop throw of 150\arcsec \, along
declination (PA=0). Two fields were observed to make a mosaic
image covering a 150\arcsec$\times$180\arcsec \, area.  Uranus was
used for flux calibration, which yielded a flux conversion factor
(FCF) of 244 JyV$^{-1}$ (per beam) with 15\% uncertainty at 850
$\rm \mu m$ and 352 JyV$^{-1}$ with 30\% uncertainty at 450 $\rm
\mu m$, respectively. Pointing was checked regularly with G34.3.
The weather was steadily good, with a CSO 225 GHz $\tau$ = 0.1
according to the JCMT water vapor monitor.

A series of molecular lines with different optical depths
including the optically thick lines HCN (3$-$2), HCO$^+$ (3$-$2),
CO (3$-$2) and optically thin lines H$^{13}$CO$^+$ (3$-$2) and
C$^{17}$O (2$-$1), were observed toward the detected SCUBA cores
in May, 2004 using the Hererodyne receivers A3 and B3 at the JCMT.
The receiver temperature, system temperature and telescope beam
size were approximately 90 K (200 K), 300--400 K (500--600 K) and
20\arcsec\ (14\arcsec) for A3 (B3). The measured main beam
efficiencies $\eta_{\rm mb}$ were 0.69 for A3 and 0.63 for B3,
with 15\% uncertainty.  A digital autocorrelation spectrometer
(DAS) with 250 MHz bandwidth was employed, which provided a
velocity resolution of 0.25 km/s.  Pointing was checked regularly
using G34.3.  Position switching mode was adopted with a reference
position of (800\arcsec , 800\arcsec).  A test observation
indicated that the reference position was free of emission at the
relevant frequencies.

HCN (3$-$2) was sampled at five points along both RA and DEC with
10\arcsec\ spacing centered at the SCUBA core. HCO$^+$ (3$-$2) and
C$^{17}$O (2$-$1) were observed in a 3$\times$3 grid. CO(3-2) was
observed at 9 points along R.A. and DEC. with 7\arcsec \ spacing.
H$^{13}$CO$^+$ (3$-$2) was measured at the center of the SCUBA
core. The integration time was 2 minutes each for CO (3-2), 5
minutes each for the HCN (3$-$2), HCO$^+$ (3$-$2) and C$^{17}$O
(2$-$1) lines and 30 minutes for H$^{13}$CO$^+$ (3$-$2).

\section{Results and discussion\label{obs_result}}
\subsection{The dust core\label{dust}}

The SCUBA images at 850 $\rm \mu m$ and 450 $\rm \mu m$ of the
NH$_3$ region 18354-0649 are presented as contours in Fig.1. For
comparison we have also obtained data from the archives of VLA,
IRAS, Spitzer and Midcourse Space Experiment (MSX). The Spitzer
IRAC 8\,$\mu$m and MSX A band, E band (Egan et al. 1998) maps are
presented as the background in Fig. 1b, 1c and 1d. The SCUBA dust
emission is concentrated in two sub-millimeter cores. The northern
core corresponds to the HII region G25.4NW which is detected in
the Spitzer and MSX maps as well as
    in the VLA 6 cm radio continuum map (Lester et al. 1985).  The
    southern core is coincident with the NH$_3$ peak, which is not
    associated with any detectable mid-IR and radio continuum emission.
    This core is therefore named as JCMT 18354$-$0649S.

The SCUBA maps show that the southern core is more compact than
the northern core. At 450 $\rm \mu m$, the northern core is
resolved, but the southern core is still barely resolved by the
8\arcsec \, beam. Follow up spectral line observations using HCN
(3$-$2), HCO$^+$ (3$-$2) and C$^{17}$O (2$-$1) show that both
cores have the same radial velocity $V_{\rm
  LSR}$ of 95 km s$^{-1}$, which is quite different from that of the
IRAS source IRAS 18355$-$0650 ($V_{\rm LSR}$ = 65 km s$^{-1}$).
This suggests that these two SCUBA cores are in the same giant
molecular cloud, but IRAS 18355$-$0650 is in a different
(foreground) molecular cloud.  Using the rotation curve of the
Galaxy (e.g. Wouterloot \& Brand 1989), the kinetic distance to
the SCUBA cores is estimated to be 5.7 or 9.6 kpc depending on
whether the source is on the near or far side of the spiral arm.
In this letter, we assume a distance of 5.7 kpc, and all the
calculations could be easily scaled to a different distance. Our
discussion will also focus on the southern core JCMT 18354-0649S.

With a photometric aperture of 25\arcsec\, the total
sub-millimeter flux is $6.38 \pm 0.96$ Jy and $41.8 \pm 12.5 $ Jy
at 850 and 450 $\mu$m for the core JCMT 18354-0649S.\footnote{
Using the formula in Seaquist et al.(2004), we estimate that the
CO(3-2) line contributes to less than 10\% of SCUBA 850 $\mu$m
fluxes at the peak position. This is within the uncertainty in the
850 $\mu$m fluxes and thus no correction was made.}  The dust
temperature T$_{\rm d}$ was estimated to be $T_{\rm d} = 14.4 $ K
according to a grey body fitting with a dust emissivity index
$\beta$=2 (T$_{\rm d}$=25.2 K if $\beta$=1.5), but it could range
from 11.0$-$29.0 K considering the uncertainty in the SCUBA fluxes
and the dust emissivity index. Our result indicates that JCMT
18354-0649S is slightly warmer than those molecular clouds
($T_{\rm d} < 15$K) without 100 $\mu$m emission (Clark et al.
1991), and is similar to, or warmer than the sourceless cores
detected by Garay et al. (2004) (upper limits from 15 to 17 K for
$\beta$=2).

The total dust/gas mass of JCMT 18354$-$0649S is calculated using $M =
 S_\nu D^2 /\kappa _\nu B_\nu(T_{\rm d})$, where $S_\nu$ is the flux
 at frequency $\nu$, $D$ is the distance (5.7 kpc), $B_\nu(T_{\rm d})$
 is the Planck function, and $\kappa _\nu$ is the dust opacity per
 unit gas/dust mass.  Using the dust opacity $\kappa _{850} = 0.02
 \,{\rm cm^2 g^{-1}}$ and $\kappa _{450} = 0.07 \,{\rm cm^2 g^{-1}}$
 calculated by Ossenkopf \& Henning (1994), and assuming a dust temperature
 $T_{\rm d}$ of 20 K, a dust to gas ratio of 0.01, we derived a total
 mass of $M$ = 910 $\,\rm M_\sun$~and 820$\,\rm M_\sun$ from the 850
 $\rm \mu m$ and 450 $\rm \mu m$ flux respectively.
The core size is 12\arcsec $\times$9\arcsec which is de-convolved with
 a 8\arcsec \ beam from the measured FWHM value (14\arcsec
 $\times$12\arcsec). Taking 10\arcsec \ as the core diameter and using
 the mass obtained with the 450 $\rm \mu m$ at $T_{\rm d}=20$ K, the
 average gas/dust density is 1.1$\times$10$^6$ cm$^{-3}$ in the
 central region with a radius of 5\arcsec, corresponding to 0.14 pc.
 These results indicate that JCMT 18354$-$0649S is a cold and high
 density core.

\subsection{Inflow and outflow motions in the southern core}

The spectra of HCN (3$-$2) and HCO$^+$ (3$-$2) are shown in Fig.
2a and Fig. 2b respectively. These data indicate that the extent
of the dense core traced by HCN (3$-$2) or HCO$^+$ (3$-$2) is
within the central $\sim 8$\arcsec \ after deconvolution, slightly
more extended than the dust core seen at SCUBA 450 $\mu$m map.
This could be an optical depth effect, since HCN or HCO$^+$(3--2)
are optically thick and would have more contribution from the
envelope surrounding the dust core.

    The profiles of H$^{13}$CO$^+$ (3$-$2) and C$^{17}$O (2$-$1) at the
    center position are presented in Fig. 2c, together with the HCN
    (3$-$2) and HCO$^+$ (3$-$2) lines for comparison.
    The most striking feature is the classical  ``blue profile'', a line
    asymmetry with the peak skewed to the blue side in the optically
    thick HCN/HCO$^+$ (3$-$2) lines, while the optically
    thin lines  H$^{13}$CO$^+$ (3$-$2) and C$^{17}$O (2$-$1) both peak at the
    absorption dip of the optically thick lines. Such feature is a
    good
    indication of inflow motions in this massive SCUBA core
    (Wu \& Evans 2003; Zhang \& Ho 1997; Zhang et al. 1998; Zhou 1999). To further
quantify the characteristics of the blue
    profile, we have calculated the parameter $\delta V$, the
    observed distribution of velocity differences defined
    by Mardones et al. (1997), derived from optically thick and thin
    lines, $\delta V = (V_{\rm thick}-
    V_{\rm thin})/\Delta V_{\rm thin}$.
    We have two optically thin  and two optically thick lines.
    Thus two peak ratios $T^*_{\rm A}({\rm B})/T^*_{\rm A}({\rm R})$ and
    four $\delta V$ were derived (Table 1). From Table 1
    we can see that the ratios $T^*_{\rm A}({\rm B})/T^*_{\rm A}({\rm R})$ from the HCN
    (3$-$2) and HCO$^+$ (3$-$2) are greater than 1, and all line pairs have  $\delta V < -0.25$, in good
    agreement with the characteristics of a collapsing core
    (Mardones et al. 1997).

For a single source, the blue profile can also be produced by
 rotation.  Taking the radius of 8\arcsec \ and a rotation velocity of
 5 kms$^{-1}$ at this radius, the dynamical mass is estimated to be
 $1300 $ M$_\sun$, which is consistent with that derived from the
 SCUBA data. However, if the blue profiles in the core JCMT
 18354$-$0649S are caused by rotation, we should also see red
 profiles. But this was not found in our spectral line maps. In
 addition, a rotating core should manifest itself with its unique
 kinematic features such as the rotation pattern in a
 position-velocity diagram, or the change of intensity ratio between
 the blue and red peaks across the rotation axis
 (Ho \& Hashick 1986). High resolution interferometer data would be
 crucial to distinguishing between the rotation and infall scenario
 for this core. With the single dish data currently available, we did
 not find any kinematic evidence for rotation.

  We have fitted the central spectra of HCN (3$-$2) and HCO$^+$
    (3$-$2) using the analytic model for collapsing clouds provided by
    Myers et al.(1996).  Without taking rotation effect into account,
    using the equation of radiative transfer and assuming two parallel
    components of equal temperature and velocity dispersion in a
    collapsing cloud, the model matches a wide range of line profiles
    in the candidate infall regions and provides a sensible estimate
    of $V_{\rm in}$, the characteristic inward speed of the gas
    forming the line.  Figs.2d and
    2e present the plots of the two line fittings
    respectively.  The model parameters defined in Myers et al. (1996)
    that best fit the two line profiles with the least square method
    are: $\tau_0 = 9.5$, $V_{\rm in} = 0.3 \,{\rm km/s}$, $T_{\rm k} =
    17.4 \,{\rm K}$, $\sigma = 1.8 \,{\rm km/s}$ for HCN (3$-$2); and
    $\tau_0 = 9.0$, $V_{\rm in} = 0.25 \,{\rm km/s}$, $T_{\rm k} =
    16.7 \,{\rm K}$, $\sigma = 1 \,{\rm km/s}$ for HCO$^+$ (3$-$2).
    The model fitted $T_{\rm k}$ agrees well with the value (17.5 K)
    that derived from the NH$_3$ data (Wu et al. 2005). The fitted
    $\tau_0$, $V_{\rm in}$, $T_{\rm k}$ and $\sigma = 1.8$ are much
    larger than those of low mass cores (Myers et al. 1996), which is
    consistent with the massive nature of the core. The reasonably
    good fit suggests that this model can be used for massive cores as
    well.  The ``kinematic'' mass infall rate, calculated using the
    formula $dM/dt= 4 \pi R_{\rm in}^2 m n V_{\rm in}$
    (Myers et al. 1996), is estimated to be 3.4$\times 10^{-3}\,
    M_{\odot}$ yr$^{-1}$. Here a radius of 8\arcsec \ was adopted for the
    infall asymmetry zone, and the gas density was assumed as the
    corresponding mean value of 2.7$\times$10$^5$ cm$^{-3}$.
    For comparison, the ``gravitational'' rate, defined
    in Shu (1977), is $1.4\times 10^{-3}$ M$_\sun$ yr$^{-1}$ for
    JCMT 18354--0649S. In the low mass cores, the ``kinematic'' mass
    infall rate agrees with the gravitational rate within a factor of
    2 (Myers et al. 1996). In our case, the ratio of these two values
    is at the higher end, possibly due to the different ISM
    environment in high mass star forming sites, e.g. higher external
    pressure.

    The HCN (3$-$2) profile at the center of the sub-millimeter core
    is remarkably broad.  As shown in Fig. 2,
    the full width of the HCN (3$-$2) line is 30 km/s.  Such a broad
    wing is a strong indication of outflow motions in the core.  It is
    larger than the total width (about 26 km/s) of the outflow of the
    high mass protostellar object near IRAS 23385+6053
    (Molinari et al. 1998). The extent of outflows of the core JCMT
    18354$-$0649S traced by HCN should be within the central
    10\arcsec, as the wing is not prominently broad at 10\arcsec \
    offset from the center. We have made follow-up observations at the
    CO (3-2) transition and the outflow wing is as broad as 38 km/s,
    from 75 to 113 km/s.  From a P-V diagram analysis, the extend of the
    outflow is $\sim 8$\arcsec  (after deconvolve with a $14''$ beam).
    Assuming an incline angle of 60 degree, the derived outflow
    timescale is 6600 yr.  The existence of outflows is consistent
    with the infall scenario, as outflow is an expected by-product of
    accretion (e.g. Churchwell 2002).  The driving source of
    outflow could be so deeply embedded that no Spitzer IRAC, MSX or
    IRAS emissions are detected.

\subsection{HMPOs and UCHII regions}

SCUBA core JCMT 18354$-$0649S is a typical cold massive core with
more than 800 $M_\odot$. As discussed above, evidence presented in
this $\sim 0.3$ pc region for both inflow and outflow motions
suggest that it might be in a hot core phase (Churchwell 2002)
--- the precursor of UC HII regions (PUCHs), which contains a
rapidly accreting massive protostar accompanied by massive bipolar
outflows. However, JCMT 18354$-$0649S is different from a normal
hot core. The average temperature is $\sim 20$ K, and there are no
MIR and IRAS sources detected in this region. This core may be at
a stage earlier than a normal hot core, and is embedded in large
amounts of cold dust.

We calculated the extinction of the core.  Taking the neutral
hydrogen to $E(B-V)$ ratio as 5.9$\times 10^{21}$ cm$^{-2}$
mag$^{-1}$ (Bohlin et al. 1978) and the interstellar average value
3.1 for $R$, we can derive the optical extinction $A_{\rm V}$ as $
3.1/(5.9\times 10^{21}/2)N({\rm H}_{2})$.  The 4.49 Jy/beam
850$\mu$m peak flux density leads to a $N({\rm H}_{2})=6.0\times
10^{23}$ cm$^{-2}$ . Therefore the optical extinction $A_{\rm V}$
is approximately 630 in the core JCMT 18354$-$0649S, much larger
than that in the low mass star forming regions (Visser et al.
2002). So it is not surprising that we could not detect any
near-IR and mid-IR emission even if a protostar does exist.

JCMT 18354$-$0649S is separated from the HII region G25.4NW with a
projected distance of 55\arcsec or 1.5 pc (Fig. 1). Both SCUBA
cores appear to be distinct isolated objects, but they have the
same radial velocity and seem to be located in the same elongated
molecular cloud detected in the $^{13}$CO (2$-$1) map (Lester et
al. 1985).  The 1.5 pc separation between these two cores is
similar to that of the two massive millimeter sources identified
by Garay et al. 2004) in an associated filament, and is in
agreement with the theoretical predictions of the fragment
separations in a filament with density of $3 \times 10^4$
cm$^{-3}$ (Fiege \& Pudritz 2000). The fact that massive cores are
often found near UC HII regions suggests that the vicinity of UC
HII region could be a good place to search for HMPOs.

We are grateful to the JCMT staff for all their assistance with the
observations. We also thank an anonymous referee for his valuable comments and
suggestions to the draft of this letter.
This project was supported by the Grant 10133020, 10128306,
10203003 of NSFC and G1999075405 of NKBRSF.

\bibliography{ref}

\clearpage
\begin{thebibliography}{}
\bibitem[Beuther et al.(2002)]{Beuther02} Beuther, H., Schilke, P., Menten, K. M., Motte, F., Sridharan, T. K. \& Wyrowski, F.  2002, \apj, 566, 945
\bibitem[Bohlin et al. (1978)]{Bohlin78} Bohlin, R. C., Savage, B. D., \& Drakes, J. F. 1978, \apj, 224, 132
\bibitem[Churchwell(2002)]{Churchwell02} Churchwell, E. 2002, ARA\&A, 40, 27
\bibitem[Clark et al. (1991)]{Clark91} Clark, F. O., Laureijs, R. J., \& Prusti, T. 1991, \apj, 731m 602
\bibitem[Egan et al.(1998)]{Egan98} Egan, M. P., Shipman, R. F., Price, S. D., Carey, S. J. \& Clark, F. O. 1998,  \apj, 494. L199
\bibitem[Fiege and Pudritz(2000)]{Fiege00} Fiege, J. D., \& Pudritz, R. E. 2000, MNRAS, 311, 105
\bibitem[Garay et al. (2004)]{Garay04} Garay, G., Faundez, S., Mardones, D., Bronfman, L., Chini, R., \& Nyman, L. 2004, \apj, 610, 313
\bibitem[Ho and Haschick (1986)]{Ho86} Ho, P. T. P., \& Haschick, A. D. 1986, \apj, 304, 501
\bibitem[Lester et al. (1985)]{Lester85} Lester, D. F., Dinerstein, H. L., Werner, M. W., Harvey, P. M., Evans, N. J., \& Brown, R. L. 1985, \apj, 296, 565
\bibitem[Mardones et al. (1997)]{Mardones97} Mardones, D., Myers, P. C., Tafalla, M., Wilner, D. J., Bachiller, R., \& Garay, G. 1997, \apj, 489, 719
\bibitem[Molinari et al.(2002)]{Molinari96} Molinari, S., Testi, L., Rodriguez, L. F., \& Zhang, Q. 2002, \apj, 570, 758
\bibitem[Molinari et al.(1998)]{Molinari98} Molinari, S., Testi, L., Brand, J., Cesaroni, R., \& Palla, F. 1998, \apj, 505, 139
\bibitem[Molinari et al.(1996)]{Molinari96} Molinari, S., Brand, J., Cesaroni, R. \& Palla, F.  1996, \aap, 308, 573
\bibitem[Myers et al.(1996)]{Myers96} Myers, P. C., Mardones, D., Tafalla, M., Williams, J. P., \& Wilner, D. J.  1996, \apj, 465, 133
\bibitem[Ossenkopf and Henning(1994)]{Ossenkopf94}Ossenkopf, V. \& Henning, T. 1994, A\&A, 291, 943
\bibitem[Seaquist et al.(2004)]{Seaquist04} Seaquist, E., Yao, L., Dunne, L., \& Cameron, H. 2004, MNRAS, 349, 1428
\bibitem[Shu(1977)]{Shu77} Shu, F. H. 1977, \apj, 214, 488
\bibitem[Sridharan et al.(2002)]{Sridharan02} Sridharan, T. K., Beuther, H., Schike, P., Menten, K. M., \& Wyrowski, F. 2002, \apj, 566, 931
\bibitem[Visser et al.(1996)]{Visser96} Visser, A. E., Richer, J. S. \& Chandler, C. J. 2002, AJ, 124, 2756
\bibitem[Wouterloot and Brand(1989)]{Wouterloot} Wouterloot, J. G. A. \& Brand, J  1989, \aaps, 80, 149
\bibitem[Wu and Evans(2003)]{Wu03} Wu, Y., \& Evans, N. J. 2003, \apj, 592, L79
\bibitem[Wu et al.(2005)]{Wu05} Wu, Y., Zhang, Q., Yu, W., Miller, M., Mao, R., Sun, K., \& Wang, Y. 2005, submitted to A\&A
\bibitem[Zhang and Ho(1997)]{Zhang97} Zhang, Q., \& Ho, P. T. P.  1997, \apj, 488, 241
\bibitem[Zhang et al.(1998)]{Zhang98} Zhang, Q., Ho, P. T. P., \& Ohashi, N. 1998, \apj, 494, 636
\bibitem[Zhou (1999)]{Zhou99} Zhou, S. Low mass star formation, ed. W. F. Wall, A. Carraminana, \& L. Carrasco, 199

\end{thebibliography}
\clearpage

\begin{figure}\epsscale{.6}\plotone{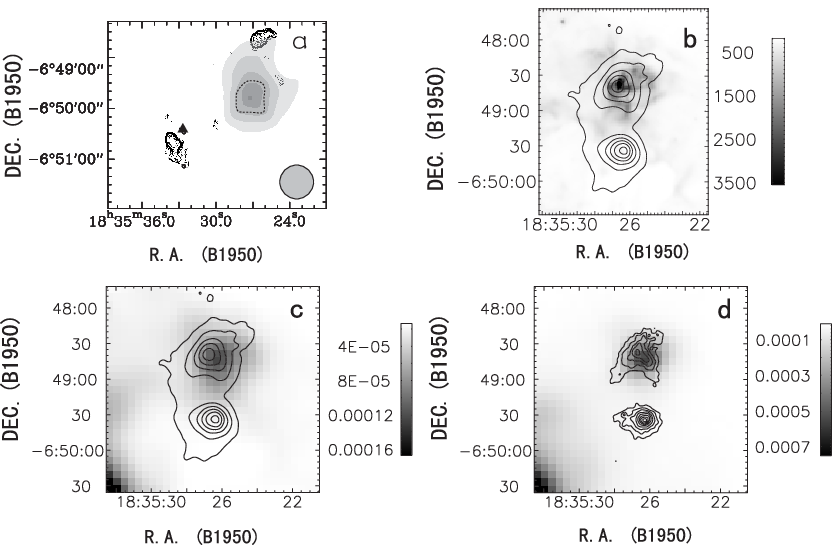}
\caption{  a) NH$_3$ (1,1) (grey scale) and (3,3) (dashed line)
overlaid on the VLA 6 cm map (solid line) of G25.4$-$0.2 (Lester
et al. 1985). The northwest and southeast clumpes are the VLA 6 cm
compacts of G25.4NW and G25.4SE, respectively (Laster et al.
1985). The solid triangle denotes IRAS 18355-0650, which is the
reference position of the NH$_3$ map. The beam size of the NH$_3$
observations is indicated in the lower right side of the panel (Wu
et al. 2005). b) SCUBA 850 $\rm \mu m$ contours on IRAC 8 $\rm \mu
m$ band image. The 8 $\rm \mu m$ flux density scale is in $\rm M
Jy$ $ sr^{-1} $. Contour levels at 850 $\rm \mu m$ are from 0.71
(10 $\sigma$) to 4.25 by 0.71 Jy/beam. c) SCUBA 850 $\rm \mu m$
contours on MSX A band image ($6.8-10.8 {\rm \mu m}$). Contour
levels are the same as in b). d) SCUBA 450 $\rm \mu m$ contours on
MSX E band image ($18.2-25.1 {\rm \mu m}$). Contour levels are
from 2.9 (5$\sigma$) to 17.7 by 2.1 Jy/beam. The MSX flux density
is in $\rm W m^{-2} sr^{-1}$. \label{fig_source}}
\end{figure}

\clearpage

\begin{figure}
\epsscale{0.53} \plotone{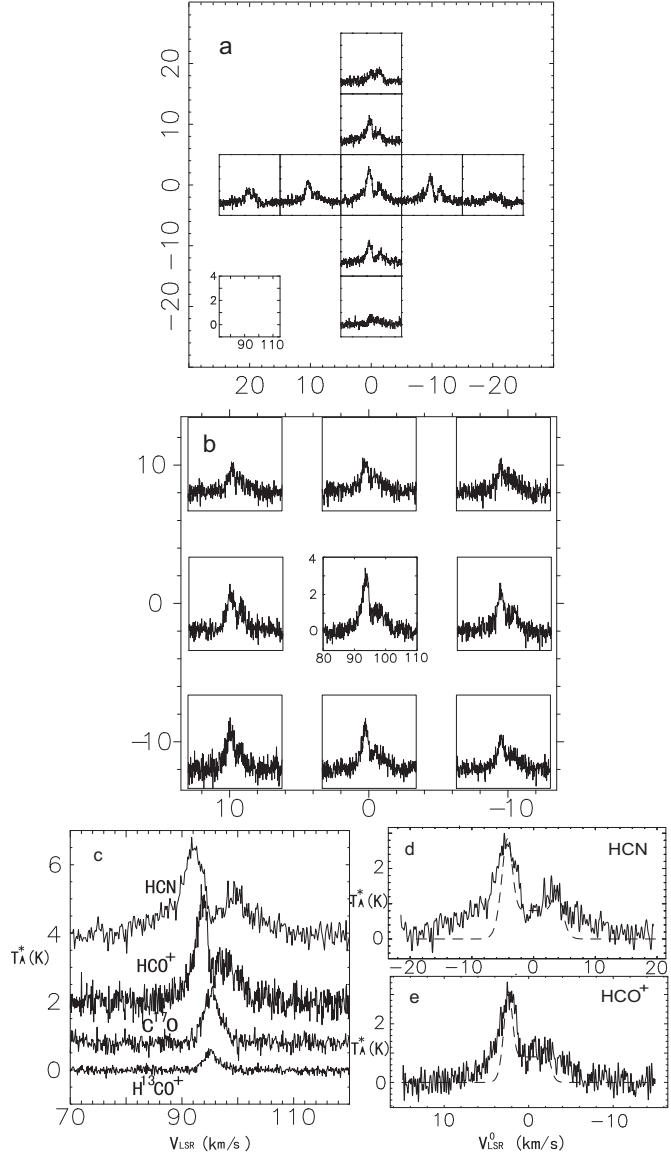} \caption{ Spectral lines of the
SCUBA core JCMT 18354$-$0649S. a) HCN (3$-$2) grid with 10\arcsec
\ spacing. X and Y axis are the R.A. and DEC. offset in arcsec,
respectively. b) HCO$^+$ (3$-$2) grid with 10\arcsec \ spacing. X
and Y axis are the same as a). c) HCN (3$-$2), HCO$^+$ (3$-$2),
C$^{17}$O (2$-$1) and H$^{13}$CO$^+$ (3$-$2) lines at the central
position. d) and e) HCN and HCO$^+$ (3$-$2) line (solid line) and
model fitting (dashed line) in zero velocity frame (indicated as
 V$_{LSR}^0$).
\label{fig_grid_hcnhco}}
\end{figure}

\clearpage

\begin{table}
    \begin{flushleft}
    \caption{Parameters to identify collapse\label{tab_collapse}}

    \begin{tabular}{cc|cc}
    \tableline
    \multicolumn{2}{c|}{\parbox[t]{20mm}{\centering {Thick line}}}&
    \multicolumn{2}{c}{\parbox[t]{20mm}{\centering {Thin line}}}    \\
    \tableline
    Molecular species & Ratio & Molecular species & $\delta V$ (km/s) \\
    \tableline
    HCO$^+$   (3$-$2) &  2.11 & C$^{17}$O (2$-$1)      & $-$0.46 \\
    \cline{3-4}
                      &       & H$^{13}$CO$^+$ (3$-$2) & $-$0.48 \\
    \tableline
    HCN       (3$-$2) & 2.26  & C$^{17}$O (2$-$1)      &  $-$0.50\\
    \cline{3-4}
                      &       & H$^{13}$CO$^+$ (3$-$2) & $-$0.52 \\
    \tableline
    \end{tabular}
    \end{flushleft}
\end{table}

\end{document}